\documentstyle[epsf]{article} 

\title{Magnetization processes
       in quantum spin chains 
       \protect\\
       with regularly alternating intersite interactions}

\author{Oleg Derzhko$^{\dagger,\ddagger}$\\
   \small $^{\dagger}$Institute for Condensed Matter Physics\\ 
   \small 1 Svientsitskii Street, L'viv--11, 79011, Ukraine\\
   \small $^{\ddagger}$Chair of Theoretical Physics,
          Ivan Franko National University in L'viv\\
   \small 12 Drahomanov Street, L'viv--5, 79005, Ukraine}        

\date{\today}

\begin{document}

\renewcommand\baselinestretch {1,5}
\large\normalsize

\maketitle

\begin{abstract}
We consider the dependence magnetization vs. field at zero temperature 
for the spin--$\frac{1}{2}$ chains 
in which the intersite interactions regularly vary from site to site 
with a period $p$.
In the limiting case 
when the smallest value of the intersite interactions tends to zero
the chain splits into noninteracting identical fragments of $p$ sites
and the dependence magnetization vs. field can be examined rigorously.
We demonstrate explicitly 
the appearance of plateaus in such dependence  
and discuss a presence of the magnetization values $m$
predicted by the condition 
$p\left(\frac{1}{2}-m\right)={\mbox{integer}}$ \cite{001}. 
We comment on 
the influence of the anisotropy in the interspin interaction 
on the magnetization profiles.
Finally,
we show 
how the case 
of nonzero smallest value of the intersite interactions
can be considered.  
\end{abstract}

\vspace{1cm}


\vspace{1cm}


\vspace{1cm}

\noindent
{\bf Postal address:}

\noindent
Dr. Oleg Derzhko\\
Institute for Condensed Matter Physics\\
1 Svientsitskii Street, L'viv--11, 79011, Ukraine\\
Tel: +380 322 76 19 78, +380 322 70 74 39\\
Fax: +380 322 76 19 78\\
E-mail: derzhko@icmp.lviv.ua

\clearpage

\renewcommand\baselinestretch {1,8}
\large\normalsize

The theoretical study of (quantum) spin chains 
attracts much attention during last years. 
On the one hand,
a number of quasi--one--dimensional magnetic compounds,
the properties of  
which can be reasonably described 
by the one--dimensional quantum spin models, 
becomes available.
On the other hand, 
quantum spin chains should exhibit various interesting properties
the examining of which 
is of great importance from the academic point of view. 
Thus, 
the analysis of the magnetization processes 
at low temperatures
may yield 
a step--like dependence magnetization vs. field. 
The latter problem has received a lot of interest  
in a numerous theoretical, numerical and experimental papers
concerning a variety of spin chains and ladders 
\cite{001,002,003}.

In what follows we discuss one mechanism
which generates a step--like dependence 
of the magnetization vs. field,
that is 
a regular alternation of the intersite interactions.
Namely,
we consider a chain of $N\to\infty$ spins $\frac{1}{2}$ 
governed by the Hamiltonian
\begin{eqnarray}
\label{001}
H=-h\sum_{n=1}^Ns_n^z
+\sum_{n=1}^N
\left(
J_n^{x}s_n^xs_{n+1}^x+J_n^{y}s_n^ys_{n+1}^y+J_n^{z}s_n^zs_{n+1}^z
\right)
\end{eqnarray}
assuming 
that the intersite (antiferromagnetic) interactions 
$J_n^{\alpha}$ ($\ge 0$)
vary regularly from site to site with a period $p$,
i.e. a sequence of parameters is
$\{J_1^x,J_1^y,J_1^z,\ldots,J_p^x,J_p^y,J_p^z,
J_1^x,J_1^y,J_1^z,\ldots\}$.
If $J_n^x=J_n^y=J_n^z=J_n$ 
Eq. (\ref{001}) 
is the Hamiltonian of the isotropic Heisenberg ($XXX$) chain,
if $J_n^x,J_n^y\ne 0$, $J_n^z=0$ 
Eq. (\ref{001}) 
corresponds to the anisotropic $XY$ chain.
For the latter chain one can differ two limiting cases,
namely,
i)
$J_n^x=J_n^y=J_n$ --- the isotropic $XY$ ($XX$) chain
and
ii)
$J_n^x=J_n$, $J_n^y=0$ --- the Ising chain.
Besides, the Hamiltonian (\ref{001}) contains 
a uniform external field $h$ directed along $z$ axis
(that is called the transverse field for $XY$ chains).

We shall be interested in the magnetization per site
$m=\frac{1}{N}\sum_{n=1}^N\langle s_n^z\rangle$
(the angle brackets denote the thermodynamical canonical average)
or more precisely  
in the dependence $m$ vs. $h$ at zero temperature. 
Oshikawa {\it et al.} \cite{001}
using the Lieb--Schultz--Mattis theorem and the bosonization techniques 
for a general quantum spin $s$ chain
with axial symmetry 
argue that the magnetization obeys
$p\left(s-m\right)={\mbox{integer}}$,
i.e. in the case $s=\frac{1}{2}$
the possible values of magnetization are
$m=\frac{p-2k}{2p}$,
$k=0,1,2,\ldots,p$. 
A lot of work has been done to check the above mentioned criteria 
using various approximate analytical approaches and numerical techniques.
On the other hand,
for the $XX$ chain in a transverse field
the magnetization $m$ can be calculated rigorously. 
The result at zero temperature reads 
\begin{eqnarray}
\label{002}
m=\frac{1}{2}\int_{-\infty}^{\infty}{\mbox{d}}E
\rho(E)(2\theta(E)-1),
\end{eqnarray} 
where $\rho(E)$ is the density of states 
of the Jordan--Wigner fermions 
which is known explicitly for any finite period $p$ \cite{004}
and $\theta(E)$ is the Heaviside step function.
Regular nonuniformity leads 
to a splitting of the fermion band of the uniform chain 
into several subbands 
that, in its turn, 
immediately leads to plateaus in the dependence $m$ vs. $h$
as follows from the above formula (\ref{002}) for $m$.
The values of the characteristic fields 
at which plateaus start and end up 
are the solutions of two algebraic equations of $p$th order,
whereas the possible values of $m$ 
are connected with 
the possible differences in the numbers of subbands 
at $E<0$ and at $E>0$ (see (\ref{002})). 
One more related work 
concerning the regularly alternating transverse Ising chain 
has been reported in Ref. \cite{005}.
Contrary to the transverse $XX$ chain, 
the regular alternation of exchange interactions 
for the transverse Ising chain 
does not lead to plateaus in the dependence 
$m$ vs. $h$.

In what follows we consider in some detail a particular case 
of the regularly alternating spin--$\frac{1}{2}$ chain (\ref{001})
when the smallest value of the intersite interactions equals to zero.
Without a loss of generality we may put 
$J_p^{\alpha}=0$.
In such limiting case a simple picture 
for explanation the zero temperature magnetization profiles 
emerges.
Really, in this limit the chain consists 
of the noninteracting clusters 
every one of which contains $p$ sites. 
The magnetization of the chain per site $m$  
follows from the magnetization of the cluster $M_p$
after dividing by $p$,
and $M_p=\langle{\mbox{GS}}\vert S_p\vert{\mbox{GS}}\rangle$,
where
$S_p=s_1^z+\ldots+s_p^z$
and
$\vert{\mbox{GS}}\rangle$
is the ground state eigenvector of the cluster Hamiltonian.
The appearance of plateaus 
arises due to the change of the ground state 
with varying of the field. 
Such a viewpoint is known as the strong--coupling approach. 
It was exploited in a number of papers 
devoted to the spin chains 
with a periodic modulation of the intersite interactions 
and the spin ladders
\cite{003}.
We discuss the strong--coupling limit  
to get a better understanding  
of the obtained earlier rigorous results 
for the transverse $XX$ and transverse Ising chains 
by means of the continued fraction approach \cite{004,005} 
as well as 
to discuss the influence of the anisotropy in spin interaction 
on the zero temperature magnetization profiles.
On the other hand, 
demonstrating how does the strong--coupling approach work 
in the exactly solvable case 
we can reveal a region of validity of this approximate method.

We start with the regularly alternating $XX$ chain with $p=2$. 
Assuming $J_2=0$ one splits the chain into noninteracting clusters 
containing two sites. The Hamiltonian of the cluster reads
\begin{eqnarray}
\label{003}
H_2=-h\left(s_1^z+s_2^z\right)
+J_1\left(s_1^xs_2^x+s_1^ys_2^y\right).
\end{eqnarray}
The eigenvalues of (\ref{003}) are
\begin{eqnarray}
\label{004}
E_1=-\frac{1}{2}J_1,
\;\;\;
E_2=-h,
\;\;\;
E_3=h,
\;\;\;
E_4=\frac{1}{2}J_1;
\end{eqnarray}
the corresponding eigenvectors are
\begin{eqnarray}
\label{005}
\vert 1\rangle=\frac{1}{\sqrt{2}}\left(
\vert\uparrow_1\downarrow_2\rangle-
\vert\downarrow_1\uparrow_2\rangle
\right),
\;\;\;
\vert 2\rangle=\vert\uparrow_1\uparrow_2\rangle,
\;\;\;
\vert 3\rangle=\vert\downarrow_1\downarrow_2\rangle,
\;\;\;
\vert 4\rangle=\frac{1}{\sqrt{2}}\left(
\vert\uparrow_1\downarrow_2\rangle+
\vert\downarrow_1\uparrow_2\rangle
\right).
\end{eqnarray}
Moreover, as it follows from (\ref{005})
$\langle 1\vert S_2\vert 1\rangle=0$,
$\langle 2\vert S_2\vert 2\rangle=1$,
$\langle 3\vert S_2\vert 3\rangle=-1$,
$\langle 4\vert S_2\vert 4\rangle=0$.
For $0\le h< \frac{1}{2}J_1$
one concludes from (\ref{004}) that
$\vert{\mbox{GS}}\rangle=\vert 1\rangle$
and therefore $M_2=0$.
For $\frac{1}{2}J_1<h$
one finds that
$\vert{\mbox{GS}}\rangle=\vert 2\rangle$
and therefore $M_2=1$.
Similarly the case $h\le 0$ can be considered.
As a result one deduces 
that the magnetization curve
$m$ vs. $h$ 
should exhibit a plateau at $m=0$ 
(if $-\frac{1}{2}J_1<h<\frac{1}{2}J_1$)
and at $m=\pm\frac{1}{2}$ 
(if $h>\frac{1}{2}J_1$ and $h<-\frac{1}{2}J_1$).

Let us consider further the case $p=3$.
The cluster Hamiltonian is as follows
\begin{eqnarray}
\label{006}
H_3=-h\left(s_1^z+s_2^z+s_3^z\right)
+J_1\left(s_1^xs_2^x+s_1^ys_2^y\right)
+J_2\left(s_2^xs_3^x+s_2^ys_3^y\right).
\end{eqnarray}
The eigenvalues of the Hamiltonian (\ref{006}) 
are as follows
\begin{eqnarray}
\label{007}
E_1=-\frac{1}{2}h-\frac{1}{2}\sqrt{J_1^2+J_2^2},
\;\;\;
E_2=\frac{1}{2}h-\frac{1}{2}\sqrt{J_1^2+J_2^2},
\;\;\;
E_3=-\frac{3}{2}h,
\;\;\;
E_4=-\frac{1}{2}h,
\nonumber\\
E_5=-E_4,
\;\;\;
E_6=-E_3,
\;\;\;
E_7=-E_2,
\;\;\;
E_8=-E_1.
\end{eqnarray}
Moreover,
$\langle 3\vert S_3\vert 3\rangle=\frac{3}{2}$,
$\langle 1\vert S_3\vert 1\rangle=
\langle 4\vert S_3\vert 4\rangle=
\langle 7\vert S_3\vert 7\rangle=\frac{1}{2}$,
$\langle 2\vert S_3\vert 2\rangle=
\langle 5\vert S_3\vert 5\rangle=
\langle 8\vert S_3\vert 8\rangle=-\frac{1}{2}$,
$\langle 6\vert S_3\vert 6\rangle=-\frac{3}{2}$.
Therefore,
for 
$0< h< \frac{1}{2}\sqrt{J_1^2+J_2^2}$
(since
$\vert{\mbox{GS}}\rangle=\vert 1\rangle$
(see (\ref{007})))
one finds $M_3=\frac{1}{2}$ and $m=\frac{1}{6}$,
whereas
for $\frac{1}{2}\sqrt{J_1^2+J_2^2}<h$
(since
$\vert{\mbox{GS}}\rangle=\vert 3\rangle$)
one finds $M_3=\frac{3}{2}$ and $m=\frac{1}{2}$.
As a result 
one concludes that the dependence 
$m$ vs. $h$ exhibits plateaus 
at $m=\frac{1}{2}$
(if $\frac{1}{2}\sqrt{J_1^2+J_2^2}<h$),
at $m=\frac{1}{6}$
(if $0<h<\frac{1}{2}\sqrt{J_1^2+J_2^2}$),
at $m=-\frac{1}{6}$
(if $-\frac{1}{2}\sqrt{J_1^2+J_2^2}<h<0$),
and 
at $m=-\frac{1}{2}$
(if $h<-\frac{1}{2}\sqrt{J_1^2+J_2^2}$).

It is interesting to note 
that the relevant for zero temperature magnetization 
low--lying levels of 
the Hamiltonians (\ref{003}) and (\ref{006})
follow from the following Hamiltonians
\begin{eqnarray}
\label{008}
{\cal{H}}_2=-h{\cal{S}}^z
+\frac{1}{2}J_1\left({\left({\cal{S}}^z\right)}^2-1\right),
\;\;\;
{\cal{S}}^z=\left\{\pm 1, 0\right\}
\end{eqnarray}
and
\begin{eqnarray}
\label{009}
{\cal{H}}_3=-h{\cal{S}}^z
+\frac{1}{4}\sqrt{J_1^2+J_2^2}
\left({\left({\cal{S}}^z\right)}^2-\frac{9}{4}\right),
\;\;\;
{\cal{S}}^z=\left\{\pm\frac{3}{2}, \pm\frac{1}{2}\right\},
\end{eqnarray}
respectively.
The appearance of plateaus becomes evident from (\ref{008}) 
(or (\ref{009})) 
since, 
e.g.,
at small $h>0$ spins are fixed to ${\cal{S}}^z=0$ 
(or to ${\cal{S}}^z=\frac{1}{2}$)
and for large $h>0$ they should be in the 
${\cal{S}}^z=\frac{3}{2}$
(or ${\cal{S}}^z=1$)
state. 

We pass to the case $p=4$.
The eigenvalues of the cluster Hamiltonian are as follows
\begin{eqnarray}
\label{010}
E_1=-\frac{1}{2}\sqrt{J_2^2+\left(J_1+J_3\right)^2}=-E_{16},
\nonumber\\
E_2=-h
-\frac{1}{2\sqrt{2}}
\sqrt{J_1^2+J_2^2+J_3^2
+\sqrt{J_1^4+J_2^4+J_3^4
-2J_1^2J_3^2+2J_1^2J_2^2+2J_2^2J_3^2}}=-E_{15},
\nonumber\\
E_3=h
-\frac{1}{2\sqrt{2}}
\sqrt{J_1^2+J_2^2+J_3^2
+\sqrt{J_1^4+J_2^4+J_3^4
-2J_1^2J_3^2+2J_1^2J_2^2+2J_2^2J_3^2}}=-E_{14},
\nonumber\\
E_4=-\frac{1}{2}\sqrt{J_2^2+\left(J_1-J_3\right)^2}=-E_{13},
\nonumber\\
E_5=-h
-\frac{1}{2\sqrt{2}}
\sqrt{J_1^2+J_2^2+J_3^2
-\sqrt{J_1^4+J_2^4+J_3^4
-2J_1^2J_3^2+2J_1^2J_2^2+2J_2^2J_3^2}}=-E_{12},
\nonumber\\
E_6=h
-\frac{1}{2\sqrt{2}}
\sqrt{J_1^2+J_2^2+J_3^2
-\sqrt{J_1^4+J_2^4+J_3^4
-2J_1^2J_3^2+2J_1^2J_2^2+2J_2^2J_3^2}}=-E_{11},
\nonumber\\
E_7=-2h=-E_{10},
\;\;\;
E_8=0=E_9.
\end{eqnarray}
Usually, a chain with $p=4$ 
(e.g., if $J_1=J_2=J_3=J$)
exhibits plateaus 
at $m=0$ for $-h_1<h<h_1$ when $\vert GS\rangle=\vert 1\rangle$,
$\langle 1\vert S_4\vert 1\rangle=0$,
at $m=\frac{1}{4}$ ($m=-\frac{1}{4}$)
for $h_1<h<h_2$ ($-h_2<h<-h_1$)
when $\vert GS\rangle=\vert 2\rangle$,
$\langle 2\vert S_4\vert 2\rangle=1$
($\vert GS\rangle=\vert 3\rangle$,
$\langle 3\vert S_4\vert 3\rangle=-1$),
and 
at $m=\frac{1}{2}$ ($m=-\frac{1}{2}$)
for $h_2<h$ ($h<-h_2$)
when $\vert GS\rangle=\vert 7\rangle$,
$\langle 7\vert S_4\vert 7\rangle=2$
($\vert GS\rangle=\vert 10\rangle$,
$\langle 10\vert S_4\vert 10\rangle=-2$).
However, for special values of parameters 
$J_1$, $J_2$, $J_3$ 
not all possible values of $m$, 
i.e. $0$, $\pm\frac{1}{4}$, $\pm\frac{1}{2}$, 
are observed. 
For example, if $E_1=E_2$ in (\ref{010})
(this occurs when $J_1=J_2=J$, $J_3=0$) 
the plateau at $m=0$ disappears.
In terms of the Jordan--Wigner fermions in such a case one has 
\cite{004}
$\rho(E)=\frac{1}{4}\delta(E+h+\frac{\sqrt{2}}{2}J)
+\frac{1}{2}\delta(E+h)
+\frac{1}{4}\delta(E+h-\frac{\sqrt{2}}{2}J)$.
Therefore, 
in accordance with (\ref{002})
$m=0$ occurs exactly when $h=0$
since any small positive (negative) $h$ 
immediately yields
$m=\frac{1}{2}$
($m=-\frac{1}{2}$).

Let us turn to the transverse Ising chain.
For $p=2$ the cluster Hamiltonian eigenvalues are
\begin{eqnarray}
\label{011}
E_1=-\sqrt{h^2+\frac{1}{16}J_1^2}=-E_4,
\;\;\;
E_2=-\frac{1}{4}J_1=-E_3.
\end{eqnarray}
Note, that the ground state is $\vert 1\rangle$ for any $h$.
Moreover, 
$\langle 1\vert S_2\vert 1\rangle=0$ 
if $h=0$
and
$\langle 1\vert S_2\vert 1\rangle\to 1$ 
if $h\to \infty$.
Thus, the considered chain does not exhibit plateaus 
that is in agreement with the result 
for a general 
(without restriction $J_2=0$)
transverse Ising chain with $p=2$ reported in \cite{005}. 
In fact, the absence of plateaus in magnetization curve is not surprising 
and is conditioned by the different symmetry 
of the transverse Ising chain. 
Thus,
$\sum_n s_n^z$ for this model with arbitrary $J_2$ 
does not commute with the Hamiltonian 
(in contrast to the case of the transverse $XX$ chain)  
and hence 
$\langle GS\vert S_p\vert GS\rangle$ 
should vary continuously with changing $h$.

One can convinced himself that the formulae 
(\ref{004}), 
(\ref{007}), 
(\ref{010}),
(\ref{011}) 
are valid for ferromagnetic sign 
of all or a part of interactions 
and hence 
the described picture is not restricted to the case $J_n\ge 0$.
For the isotropic Heisenberg chain with $p=2$
with the antiferromagnetic interaction $J_1>0$
the cluster Hamiltonian eigenvalues are 
$E_1=-\frac{3}{4}J_1$,
$E_2=-h+\frac{1}{4}J_1$,
$E_3=\frac{1}{4}J_1$,
$E_4=h+\frac{1}{4}J_1$,
and therefore one concludes that 
$m=0$ for $-J_1<h<J_1$ 
and 
$m=\frac{1}{2}$
($m=-\frac{1}{2}$)
for 
$J_1<h$
($h<-J_1$).
However, for the ferromagnetic interaction $J_1<0$
one finds instead
$E_1=-h-\frac{1}{4}\vert J_1\vert$,
$E_2=-\frac{1}{4}\vert J_1\vert$,
$E_3=h-\frac{1}{4}\vert J_1\vert$,
$E_4=-\frac{3}{4}\vert J_1\vert$,
and hence the plateau at $m=0$ does not appear.

Evidently, in the considered limit $J^{\alpha}_p=0$  
an arbitrary type of the intersite interaction 
(the anisotropic $XY$ chain,
the Heisenberg--Ising ($XXZ$) chain etc.)
and the value of spin $s$
can be examined easily.
In addition to the total magnetization 
the on--site magnetization can be calculated in a similar way.
Besides, the treatment in that limit 
can be applied to the spin systems of higher dimensions.

Finally, let us discuss briefly 
how the obtained results can be used for the analysis 
beyond the limit $J_p=0$.
The transverse $XX$ chain or the transverse Ising chain 
can be studied rigorously \cite{004,005}
in contrast to the Heisenberg or more complicated chains.
Different ways to take into account 
a nonzero value of the smallest interaction 
perturbatively at certain value of $m$ 
have been elaborated \cite{003}.
Let us demonstrate how it can be done 
considering for concreteness 
the regularly alternating $XX$ chain of period 2 
in which now $J_1\gg J_2\ne 0$. 
The Hamiltonian of that system naturally splits into two parts
\begin{eqnarray}
\label{012}
H=H_0+V,
\\
H_0=\ldots-h\left(s^z_{101}+s^z_{102}\right)
+\frac{1}{2}J_1\left(s^+_{101}s^-_{102}+s^-_{101}s^+_{102}\right)-\ldots,
\nonumber\\
V=\ldots+\frac{1}{2}J_2\left(s^+_{102}s^-_{103}+s^-_{102}s^+_{103}\right)+\ldots.
\nonumber
\end{eqnarray}
Only the two lowest levels of the two--site cluster discussed above 
will be taken into account.
We assume $h\ge 0$;
the relevant states are
$\vert 1\rangle_{51}=
\frac{1}{\sqrt{2}}
\left(
\vert\uparrow_{101}\downarrow_{102}\rangle
-\vert\downarrow_{101}\uparrow_{102}\rangle
\right)$
and 
$\vert 2\rangle_{51}=
\vert\uparrow_{101}\uparrow_{102}\rangle$.
Let us introduce new spin $\frac{1}{2}$ operators $\sigma^{\alpha}_l$
which act in a following way
\begin{eqnarray}
\label{013}
\ldots,\;\;\;
\sigma^z_{51}\vert 1\rangle_{51}=-\frac{1}{2}\vert 1\rangle_{51},
\;\;\;
\sigma^+_{51}\vert 1\rangle_{51}=\vert 2\rangle_{51},
\;\;\;
\sigma^-_{51}\vert 1\rangle_{51}=0,
\nonumber\\
\sigma^z_{51}\vert 2\rangle_{51}=\frac{1}{2}\vert 2\rangle_{51},
\;\;\;
\sigma^+_{51}\vert 2\rangle_{51}=0,
\;\;\;
\sigma^-_{51}\vert 2\rangle_{51}=\vert 1\rangle_{51},
\;\;\;\ldots\;.
\end{eqnarray}
Then $H_0$ can be written approximately 
(only the lowest levels are reproduced)
as
\begin{eqnarray}
\label{014}
H_0=\sum_{l=1}^{L}
\left(
-\frac{1}{2}J_1\left(\frac{1}{2}-\sigma^z_l\right)
-h\left(\frac{1}{2}+\sigma^z_l\right)
\right)
=-\frac{1}{2}\left(h+\frac{1}{2}J_1\right)L
-\left(h-\frac{1}{2}J_1\right)\sum_{l=1}^{L}\sigma^z_l
\end{eqnarray}
with
$L=\frac{1}{2}N$.
Really,
\begin{eqnarray}
\label{015}
\left(
-\frac{1}{2}\left(h+\frac{1}{2}J_1\right)
-\left(h-\frac{1}{2}J_1\right)\sigma^z_l
\right)
\vert 1\rangle_l
=-\frac{1}{2}J_1\vert 1\rangle_l,
\nonumber\\
\left(
-\frac{1}{2}\left(h+\frac{1}{2}J_1\right)
-\left(h-\frac{1}{2}J_1\right)\sigma^z_l
\right)
\vert 2\rangle_l
=-h\vert 2\rangle_l,
\end{eqnarray}
that are just the two lowest levels of $l$th cluster.
To write the intercluster interaction, e.g.,
$V_{51,52}=\frac{1}{2}J_2
\left(s^+_{102}s^-_{103}+s^-_{102}s^+_{103}\right)$
in terms of the operators $\sigma^{\alpha}_l$
let us consider the following equations
\begin{eqnarray}
\label{016}
V_{51,52}\vert 1\rangle_{51}
=\frac{1}{2}J_2s^-_{103}\frac{1}{\sqrt{2}}\vert 2\rangle_{51}
=\frac{1}{2\sqrt{2}}J_2s^-_{103}\sigma^+_{51}\vert 1\rangle_{51},
\nonumber\\
V_{51,52}\vert 2\rangle_{51}
=\frac{1}{2}J_2s^+_{103}\vert\uparrow_{101}\downarrow_{102}\rangle
=\frac{1}{2}J_2s^+_{103}\frac{1}{\sqrt{2}}\vert 1\rangle_{51}
=\frac{1}{2\sqrt{2}}J_2s^+_{103}\sigma^-_{51}\vert 2\rangle_{51}
\end{eqnarray}
(note, that although 
$\vert\uparrow_{101}\downarrow_{102}\rangle
=\frac{1}{\sqrt{2}}\left(\vert 1\rangle_{51}+\vert 4\rangle_{51}\right)$,
we put
$\vert\uparrow_{101}\downarrow_{102}\rangle
=\frac{1}{\sqrt{2}}\vert 1\rangle_{51}$
since we take into account only 
$\vert 1\rangle_{51}$ and $\vert 2\rangle_{51}$).
Hence
\begin{eqnarray}
\label{017}
V_{51,52}
=\frac{1}{2\sqrt{2}}J_2
\left(
s^-_{103}\sigma^+_{51}
+s^+_{103}\sigma^-_{51}
\right).
\end{eqnarray}
Further,
\begin{eqnarray}
\label{018}
V_{51,52}\vert 1\rangle_{52}
=\frac{1}{2\sqrt{2}}J_2
\left(
\sigma^+_{51}
\frac{1}{\sqrt{2}}\vert 3\rangle_{52}
+
\sigma^-_{51}
\frac{1}{\sqrt{2}}\left(-\vert 2\rangle_{52}\right)\right)
=-\frac{1}{4}J_2\sigma^-_{51}\sigma^+_{52}\vert 1\rangle_{52},
\nonumber\\
V_{51,52}\vert 2\rangle_{52}
=\frac{1}{2\sqrt{2}}J_2
\sigma^+_{51}\vert\downarrow_{103}\uparrow_{104}\rangle
=-\frac{1}{4}J_2\sigma^+_{51}\sigma^-_{52}\vert 2\rangle_{52}.
\end{eqnarray}
As a result one concludes that 
\begin{eqnarray}
\label{019}
V_{51,52}
=-\frac{1}{4}J_2
\left(\sigma^+_{51}\sigma^-_{52}+\sigma^-_{51}\sigma^+_{52}\right)
\end{eqnarray}
and therefore
\begin{eqnarray}
\label{020}
V
=-\frac{1}{2}J_2
\sum_{l=1}^{L}
\left(\sigma^x_{l}\sigma^x_{l+1}
+\sigma^y_{l}\sigma^y_{l+1}\right).
\end{eqnarray}
The Hamiltonian (\ref{012}), (\ref{014}), (\ref{020})
describes the {\it uniform} 
spin--$\frac{1}{2}$ transverse $XX$ chain of $L$ spins
\begin{eqnarray}
\label{021}
H=-\frac{1}{2}\left(h+\frac{1}{2}J_1\right)L
-\left(h-\frac{1}{2}J_1\right)\sum_{l=1}^{L}\sigma^z_l
-\frac{1}{2}J_2
\sum_{l=1}^{L}
\left(\sigma^x_{l}\sigma^x_{l+1}
+\sigma^y_{l}\sigma^y_{l+1}\right).
\end{eqnarray}
Finally, acting like while deriving (\ref{017}), (\ref{019})
one finds
the relation between the operators 
$s^{\alpha}_n$
and
$\sigma^{\alpha}_l$
\begin{eqnarray}
\label{022}
\ldots,
\;\;\;
s^+_{101}=-\frac{1}{\sqrt{2}}\sigma^+_{51},
\;\;\;
s^-_{101}=-\frac{1}{\sqrt{2}}\sigma^-_{51},
\;\;\;
s^z_{101}=\frac{1}{2}\left(\frac{1}{2}+\sigma^z_{51}\right),
\nonumber\\
s^+_{102}=\frac{1}{\sqrt{2}}\sigma^+_{51},
\;\;\;
s^-_{102}=\frac{1}{\sqrt{2}}\sigma^-_{51},
\;\;\;
s^z_{102}=\frac{1}{2}\left(\frac{1}{2}+\sigma^z_{51}\right),
\;\;\;
\ldots\;.
\end{eqnarray}
Consider now the magnetization
which owing to (\ref{022}) can be written as 
$m=\frac{1}{2}\left(\frac{1}{2}
+\frac{1}{L}\sum_{l=1}^{L}\langle\sigma_l^z\rangle\right)$.
Using the well--known results 
for the uniform 
spin--$\frac{1}{2}$ transverse $XX$ chain
(see, e.g. \cite{004}) 
one concludes that 
$m=0$ at $0\le h\le\frac{1}{2}\left(J_1-J_2\right)$
and 
$m=\frac{1}{2}$ at $\frac{1}{2}\left(J_1+J_2\right)\le h$.
If $h$ increases from 
$\frac{1}{2}\left(J_1-J_2\right)$
to
$\frac{1}{2}\left(J_1+J_2\right)$
the magnetization increases 
from $0$ to $\frac{1}{2}$
as
$\frac{1}{2}-\frac{1}{2\pi}
\arcsin \sqrt{1-\frac{\left(2h-J_1\right)^2}{{J_2}^2}}$.

In Fig. 1 we plotted 
the zero temperature magnetization profiles 
for the spin--$\frac{1}{2}$ transverse $XX$ chain of period 2 
with $J_1=1+\delta$, $J_2=1-\delta$ 
for $\delta=1,\;0.8,\;0.6,\;0.4,\;0.2,\;0$
as they follow from 
1) the exact formula (\ref{002}) \cite{004}
(solid curves)
and 
2) the approximate Hamiltonian (\ref{021}) (dotted curves).
These plots demonstrate how does the strong--coupling approach work 
as $\delta$ deviates from 1.

To summarize, we have reconsidered 
the zero temperature magnetization processes 
in the regularly alternating spin--$\frac{1}{2}$ $XY$ chains 
within the frames of the strong--coupling approach.
Besides discussing the magnetization plateaus 
in terms of the spins 
rather than in terms of the Jordan--Wigner fermions 
we have demonstrated 
to what extent the strong--coupling approximation 
can reproduce the exact magnetization profiles.

\vspace{5mm}

The author is grateful to 
J.~Richter, N.~B.~Ivanov, T.~Krokhmalskii, O.~Zaburannyi
and V.~Derzhko
for discussions.
He thanks J.~Richter for hospitality in the Magdeburg University 
in summer of 2000 when the paper was completed.

\vspace{2cm}

\noindent
\Large{{\bf Figure caption}}\\
\normalsize

\vspace{0.3cm}

Fig. 1.
$m$ vs. $h$ for
the spin--$\frac{1}{2}$ transverse $XX$ chain of period 2
with $J_1=1+\delta$, $J_2=1-\delta$, 
$\delta=1$ (a),
$\delta=0.8$ (b),
$\delta=0.6$ (c),
$\delta=0.4$ (d),
$\delta=0.2$ (e),
$\delta=0$ (f)
at zero temperature.
Solid curves correspond to the exact results
following from (\ref{002}) \cite{004},
dotted curves correspond to the results 
obtained with the help of the approximate Hamiltonian (\ref{021}).

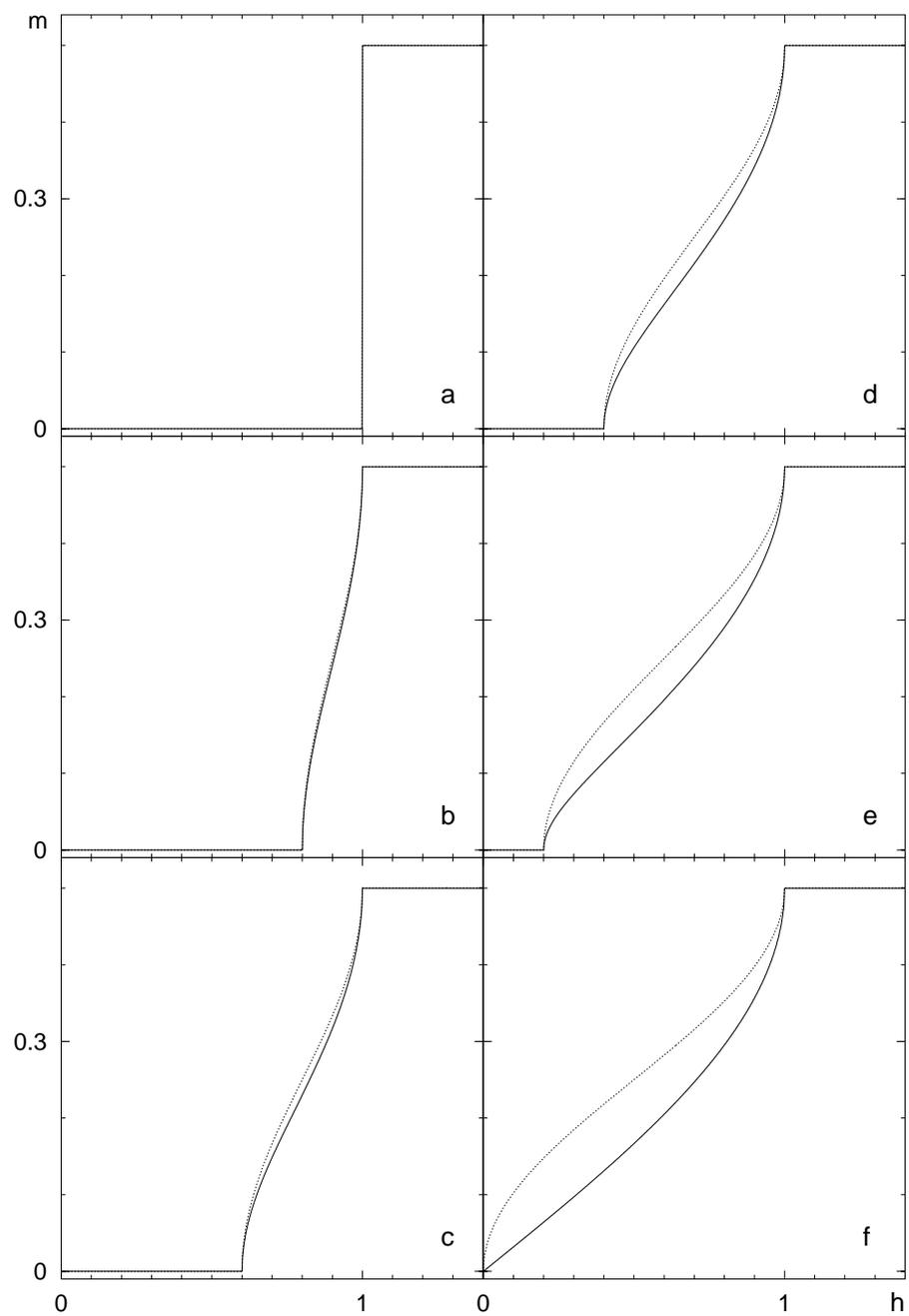
\end{document}